\documentclass[a4paper,final,notitlepage,onecolumn]{article}
\usepackage{amsmath}
\usepackage{acronym}

\input{tcilatex}

\begin{document}

\title{Observers in spacetimes with spherical and axial symmetries}
\author{Pawel Gusin*$^{1}$, Bartosz Ku\'{s}nierz**, Andrzej Radosz** \\
Wroc\l aw University of Technology,\\
*Faculty of Technology and Computer Science, Jelenia G\'{o}ra, \\
**Department of Quantum Technologies, Wroc\l aw; Poland}
\maketitle

\begin{abstract}
We introduce in the explicit form the tetrads of arbitrary observers in
spacetimes with spherical and axial symmetries. The observers confined to
the equatorial plane are parametrized by the pair of functions. We apply
this description in the analysis of the null-geodesics in the observers'
frames. The observers with the constant acceleration are distinguished.

\begin{description}
\item 
\begin{description}
\item[PACS] : 04.20.-q ; 04.90.+e
\end{description}

\item[Keywords] : tetrad, observers, null-geodesics

\item[*$^{1}$)] {\small corresponding author: pawel.gusin@pwr.edu.pl}
\end{description}
\end{abstract}

\section{Introduction}

Problem of the measurements in an observer frame is essential for any
physical theory. In physical general relativistic system the measurements
are observer-dependent so, it is important to compare their outcomesin
different frames of reference. On the other hand the measurement has a
physical meaning if it is given by the diffeomorphism invariant quantity.
Thus one need to identify such invariants. This problem has been solved long
time ago by \'{E}lie Cartan [1] for the manifolds. However the problem of
observables in general relativity have been reconsidered e.g. [2, 3, 4].

In this paper we will consider observers in space-times with fixed symmetry.
We will establish the parametrization of the frames of reference of these
observers by the pair of functions. In general the observer is determined by
a timelike normalized (local) vector field on spacetime and provides a
one-dimensional (timelike) foliation. Thus one defines the observer's frame
as the manner of the splitting of the spacetime $M$ on the time and space.
Locally this splitting is always possible and means that the plane $T_{x}M$
tangent\ to $M$ in the point $x$ has the decomposition: $T_{x}M=\mathbf{R}%
u_{\left( 0\right) }\oplus \Sigma _{p}$, where $u_{\left( 0\right) }$ is a
timelike vector and $\Sigma _{p}$ is a vector space given by the three
spacelike vectors $\{u_{\left( i\right) }\}.$ These vectors constitute a
local orthonormal basis: $\left\langle u_{\left( \alpha \right) },u_{\left(
\beta \right) }\right\rangle =\eta _{\alpha \beta }$ in the spacetime metric 
$\left\langle \cdot ,\cdot \right\rangle $. Hence the all measures are
performed with respect to this vector basis.

In the section 2 we will define the family of radial and transversal
observers in the spherically symmetric spacetimes identifying relations
between them. We will distinguish the class of observers with the constant
acceleration. They are generalizing the observers moving on the hyperbolic
trajectories in the flat Minkowski spacetime.

In the section 3 we will consider the null geodesics in the frames of the
observers with the constant acceleration. As the examples we will find
relations of these null geodesics to the observers in the Schwarzschild and
the Reissner--Nordstr\"{o}m--de Sitter spacetimes.

In the section 4 we will generalize the results of the section 2 to the
spacetimes with the axial symmetry and in the section 5 we will apply the
obtained results to the Kerr spacetime. The section 6 is devoted to
conclusions.

\section{Observers in the spherically symmetric spacetimes}

An arbitrary observer is determined by a velocity time-like vector field $u$ 
\begin{equation}
u^{2}=1.  \tag{2.1}
\end{equation}%
This vector field gives the splitting of the space-time on the time and
space. The temporal direction is given by $u_{\left( 0\right) }\equiv u$ and
the spatial directions are given by three space-like vector fields: $%
u_{\left( i\right) }$ and $i=1,2,3$. Vector fields $\left( u_{\left(
0\right) },u_{\left( i\right) }\right) \equiv \left( u_{\left( \alpha
\right) }\right) $ (where $\alpha =0,..,3$) form the local tetrad:%
\begin{equation}
u_{\left( \alpha \right) }\cdot u_{\left( \beta \right) }\equiv g_{\mu \nu
}u_{\left( \alpha \right) }^{\mu }u_{\left( \beta \right) }^{\nu }=\eta
_{\alpha \beta },  \tag{2.2}
\end{equation}%
where $\left( \eta _{\alpha \beta }\right) =diag\left( +1,-1,-1,-1\right) $.
In a general case the vector field $u$ has the four components: $\left(
u^{t},u^{r},u^{\theta },u^{\phi }\right) $ (in the spherical coordinate
system $\left( r,\theta ,\phi \right) $ ). The 10 equations (2.2) relate 16
components of the tetrad $\left( u_{\left( \alpha \right) }^{\mu }\right) $.
Hence the tetrad of the observer is determined by the six independent
parameters.

In a static spherically symmetric spacetime $M$ with a metric $g_{\mu \nu }$:%
\begin{equation}
ds^{2}=g_{0}dt^{2}-g_{1}dr^{2}-g_{2}d\theta ^{2}-g_{3}d\phi ^{2},  \tag{2.3}
\end{equation}%
where the components $g_{0}$ and $g_{1}$ depend only on the "radial"
coordinate $r$ and: $g_{2}=r^{2}$, $g_{3}=r^{2}\sin ^{2}\theta $ we will
consider a class of observers following planar trajectories. Without loss of
generality we will assume that $\theta =\pi /2$ so the vector field $u$ has
the form:%
\begin{equation}
u=\alpha \partial _{t}+\beta \partial _{r}+\gamma \partial _{\phi } 
\tag{2.4}
\end{equation}%
\ We can parametrize the vector field (2.4) by a pair of functions $\left(
q\left( r\right) ,\chi \left( r\right) \right) $ as follows:%
\begin{equation}
u=\frac{1}{\sqrt{g_{0}}}\cosh q\partial _{t}+\frac{1}{\sqrt{g_{1}}}\sinh
q\cos \chi \partial _{r}+\frac{1}{\sqrt{g_{3}}}\sinh q\sin \chi \partial
_{\phi }.  \tag{2.5}
\end{equation}%
From the equations (2.2) we get the other space-like vectors $u_{\left(
i\right) }$:%
\begin{equation}
u_{\left( 1\right) }=\frac{1}{\sqrt{g_{0}}}\sinh q\partial _{t}+\frac{1}{%
\sqrt{g_{1}}}\cosh q\cos \chi \partial _{r}+\frac{1}{\sqrt{g_{3}}}\cosh
q\sin \chi \partial _{\phi },  \tag{2.6}
\end{equation}%
\begin{equation}
u_{\left( 2\right) }=\frac{1}{\sqrt{g_{2}}}\partial _{\theta },  \tag{2.7}
\end{equation}%
\begin{equation}
u_{\left( 3\right) }=-\frac{\sin \chi }{\sqrt{g_{1}}}\partial _{r}+\frac{%
\cos \chi }{\sqrt{g_{3}}}\partial _{\phi }.  \tag{2.8}
\end{equation}%
In this way we have obtained the tetrad given by (2.5-8). The (equatorial)
trajectory of the observer (2.5) is given by an integral curve $\gamma
=\left( t\left( \lambda \right) ,r\left( \lambda \right) ,\theta \left(
\lambda \right) =\pi /2,\phi \left( \lambda \right) \right) $ of $u:$%
\begin{equation}
\frac{dt}{d\lambda }=\frac{1}{\sqrt{g_{0}}}\cosh q,\text{ \ \ }\frac{dr}{%
d\lambda }=\frac{1}{\sqrt{g_{1}}}\sinh q\cos \chi ,  \tag{2.9}
\end{equation}%
\begin{equation}
\frac{d\theta }{d\lambda }=0,\text{ \ \ }\frac{d\phi }{d\lambda }=\frac{1}{%
\sqrt{g_{3}}}\sinh q\sin \chi ,  \tag{2.10}
\end{equation}%
where $\lambda $ is a parameter of $\gamma $. An acceleration $a=a^{\mu
}\partial _{\mu }$ of the observer (2.5) is given by:%
\begin{equation}
a=\nabla _{u}u=u^{\nu }\left( \partial _{\nu }u^{\mu }+\Gamma _{\nu \rho
}^{\mu }u^{\rho }\right) \partial _{\mu }.  \tag{2.13}
\end{equation}%
The Christoffel symbols $\Gamma _{\nu \rho }^{\mu }$ for the metric (2.3)
and the components of $a$ are presented in the Appendix. Thus we get the
relations ($a^{\theta }=0$ ): 
\begin{equation}
a^{t}=\frac{\sinh q\cos \chi }{g_{0}\sqrt{g_{1}}}\frac{d}{dr}\left[ \sqrt{%
g_{0}}\cosh q\right] ,  \tag{2.14}
\end{equation}%
\begin{equation}
a^{r}=\frac{1}{2g_{1}}\frac{d}{dr}\left( \sinh ^{2}q\cos ^{2}\chi \right) +%
\frac{g_{0}^{\prime }}{2g_{1}g_{0}}\cosh ^{2}q-\frac{1}{rg_{1}}\sinh
^{2}q\sin ^{2}\chi ,  \tag{2.15}
\end{equation}%
\begin{equation}
a^{\phi }=\frac{\sinh q\cos \chi }{r^{2}\sin \theta \sqrt{g_{1}}}\frac{d}{dr}%
\left( r\sinh q\sin \chi \right) .  \tag{2.16}
\end{equation}%
Moreover the acceleration is orthogonal to the velocity: $a\cdot u=0$ (the
dot means the scalar product with respect to the metric 2.3) Thus in the
general case $a$ is the linear combination of the vectors (2.6-8): $%
a=a_{i}u_{\left( i\right) }$, where $i=1,2,3$.

There is the special class of the observers with the constant acceleration: $%
a\cdot a=\alpha ^{2}=const.$ In the basis $u_{\left( i\right) }$ this
condition has the form: $\left( a_{1}\right) ^{2}+\left( a_{2}\right)
^{2}+\left( a_{3}\right) ^{2}=\alpha ^{2}$ and $a_{i}$ are the constant
number. Because $a^{\theta }=0$ we get $a_{2}=0$ and the equations (2.14-16)
take the form:%
\begin{equation}
\cos \chi \frac{d}{dr}\left[ \sqrt{g_{0}}\cosh q\right] =a_{1}\sqrt{%
g_{0}g_{1}},  \tag{2.17}
\end{equation}%
\begin{gather}
\frac{d}{dr}\left( \sinh ^{2}q\cos ^{2}\chi \right) +\frac{g_{0}^{\prime }}{%
g_{0}}\cosh ^{2}q-\frac{2}{r}\sinh ^{2}q\sin ^{2}\chi =  \notag \\
2\sqrt{g_{1}}\left( a_{1}\cosh q\cos \chi -a_{3}\sin \chi \right) , 
\tag{2.18}
\end{gather}%
\begin{equation}
\sinh q\cos \chi \frac{d}{dr}\left( r\sinh q\sin \chi \right) =\sqrt{%
g_{1}g_{3}}\left( a_{1}\cosh q\sin \chi +a_{3}\cos \chi \right) .  \tag{2.19}
\end{equation}%
There is a special solution of these equations when $a_{1}=a_{2}=a_{3}=0$.
It has the form:%
\begin{equation}
\cosh q=E/\sqrt{g_{0}},  \tag{2.20}
\end{equation}%
\begin{equation}
\sinh q\cos \chi =\pm \sqrt{\frac{E^{2}}{g_{0}}-\frac{L^{2}\sin ^{2}\theta }{%
r^{2}}-1},  \tag{2.21}
\end{equation}%
\begin{equation}
\sinh q\sin \chi =\frac{L\sin \theta }{r}.  \tag{2.22}
\end{equation}%
This solution corresponds to the time-like geodesic of the observer given by:%
\begin{equation}
u=\frac{E}{g_{0}}\partial _{t}\pm \frac{1}{\sqrt{g_{1}}}\left[ \sqrt{\frac{%
E^{2}}{g_{0}}-\frac{L^{2}\sin ^{2}\theta }{r^{2}}-1}\right] \partial _{r}+%
\frac{L}{r^{2}}\partial _{\phi }.  \tag{2.23}
\end{equation}

Next we consider two classes of observers: the \emph{radia}l observers $\chi
=0$ and the \emph{transversal} observers, $\chi =\pi /2.$ These two classes
parametrized by the one function $q\left( r\right) $ only

\subsection{The radial observer, $\protect\chi =0$}

Let us first examine the radial observers. In this case the tetrad is:

\begin{equation}
u=\frac{1}{\sqrt{g_{0}}}\cosh q\partial _{t}+\frac{1}{\sqrt{g_{1}}}\sinh
q\partial _{r},  \tag{2.24}
\end{equation}%
\begin{equation}
u_{\left( 1\right) }=\frac{1}{\sqrt{g_{0}}}\sinh q\partial _{t}+\frac{1}{%
\sqrt{g_{1}}}\cosh q\partial _{r},  \tag{2.25}
\end{equation}%
\begin{equation}
u_{\left( 2\right) }=\frac{1}{\sqrt{g_{2}}}\partial _{\theta }\text{ , }%
u_{\left( 3\right) }=\frac{1}{\sqrt{g_{3}}}\partial _{\phi }.  \tag{2.26}
\end{equation}%
The trajectory of the this observer is given by the equations$:$%
\begin{equation}
\frac{dt}{d\lambda }=\frac{1}{\sqrt{g_{0}}}\cosh q,\text{ \ \ }\frac{dr}{%
d\lambda }=\frac{1}{\sqrt{g_{1}}}\sinh q,  \tag{2.27}
\end{equation}%
\begin{equation}
\frac{d\theta }{d\lambda }=0,\text{ \ \ }\frac{d\phi }{d\lambda }=0, 
\tag{2.28}
\end{equation}%
where $\lambda $ is a parameter of $\gamma $. For the observer with the
constant acceleration the equations (2.17-19) have the solution:%
\begin{equation}
\cosh q\left( r\right) =\frac{1}{\sqrt{g_{0}}}\left( E+\alpha \int^{r}\sqrt{%
g_{0}g_{1}}dr\right)  \tag{2.30}
\end{equation}%
\begin{equation}
\sinh q\left( r\right) =\pm \sqrt{\frac{1}{g_{0}}\left( E+\alpha \int^{r}%
\sqrt{g_{0}g_{1}}dr\right) ^{2}-1},  \tag{2.31}
\end{equation}%
where $a_{1}=\alpha $ and $a_{3}=0$. The radial observer moves on the
geodesic if $a^{t}=a^{r}=0$. This condition leads to the relation: $\cosh
q=E/\sqrt{g_{0}}$ with the vector field:%
\begin{equation}
u=\frac{E}{g_{0}}\partial _{t}+\frac{1}{\sqrt{g_{1}}}\left[ \frac{E^{2}}{%
g_{0}}-1\right] ^{1/2}\partial _{r}.  \tag{2.32}
\end{equation}

\subsection{The transversal observer, $\protect\chi =\protect\pi /2$}

In this case the tetrad has basis vectors:

\begin{equation}
u=\frac{1}{\sqrt{g_{0}}}\cosh q\partial _{t}+\frac{1}{\sqrt{g_{3}}}\sinh
q\partial _{\phi },  \tag{2.33}
\end{equation}%
\begin{equation}
u_{\left( 1\right) }=\frac{1}{\sqrt{g_{0}}}\sinh q\partial _{t}+\frac{1}{%
\sqrt{g_{3}}}\cosh q\partial _{\phi },  \tag{2.34}
\end{equation}%
\begin{equation}
u_{\left( 2\right) }=\frac{1}{\sqrt{g_{2}}}\partial _{\theta },\text{ \ }%
u_{\left( 3\right) }=-\frac{1}{\sqrt{g_{1}}}\partial _{r}.  \tag{2.35}
\end{equation}%
The trajectory of such an observer is given by the equations$:$%
\begin{equation}
\frac{dt}{d\lambda }=\frac{1}{\sqrt{g_{0}}}\cosh q,\text{ \ \ }\frac{d\phi }{%
d\lambda }=\frac{1}{\sqrt{g_{3}}}\sinh q,  \tag{2.36}
\end{equation}%
\begin{equation}
\frac{d\theta }{d\lambda }=0,\text{ \ \ }\frac{dr}{d\lambda }=0,  \tag{2.37}
\end{equation}%
where $\lambda $ is a parameter of $\gamma $. For the observer with the
constant acceleration $\alpha $ the equations (2.17-19) have the solution:%
\begin{equation}
\cosh ^{2}q\left( r\right) =2g_{0}\frac{1+ra_{3}\sqrt{g_{1}}}{%
2g_{0}-rg_{0}^{\prime }}\geq 1,  \tag{2.38}
\end{equation}%
\begin{equation}
\sinh ^{2}q\left( r\right) =2g_{0}\frac{1+ra_{3}\sqrt{g_{1}}}{%
2g_{0}-rg_{0}^{\prime }}-1,  \tag{2.39}
\end{equation}%
where $a_{1}=0$ and $\left( a_{3}\right) ^{2}=\alpha ^{2}$. From the
eq.(2.39) we obtain the allowed radius of the transversal observer:%
\begin{equation}
2g_{0}-rg_{0}^{\prime }>0  \tag{2.40}
\end{equation}%
and:%
\begin{equation}
2g_{0}a_{3}\sqrt{g_{1}}+g_{0}^{\prime }>0.  \tag{2.41}
\end{equation}%
In the case of the "free fall" $a_{3}=0$ comparing (2.38-39) and (2.21-23)
one obtains:%
\begin{equation}
rg_{0}=\frac{1}{\sqrt{2}}\frac{L}{\sqrt{1-1/E^{2}}}.  \tag{2.42}
\end{equation}%
The relations (2.40) and (2.41) for the Schwarzschild metric $%
g_{0}=1-r_{S}/r $ yield the constraint on $L$ and $E$:%
\begin{equation*}
2L^{2}\sin ^{2}\theta >r_{S}^{2}\left( 1-1/E^{2}\right)
\end{equation*}%
and the well known condition on the allowed radii of the circular orbits: $%
r>3r_{S}/2$.

\subsection{Relative velocity}

According to [5] the relative velocity $v$ between two observers determined
by $u$ and $u^{\prime }$ at the same point $x$ is:%
\begin{equation}
v=\frac{1}{g\left( u^{\prime },u\right) }u^{\prime }-u.  \tag{2.43}
\end{equation}%
In our parametrization these observers are given by two pairs of functions $%
\left( q,\chi \right) $ and $\left( q^{\prime },\chi ^{\prime }\right) $.
Hence the scalar product $g\left( u^{\prime },u\right) $ is equal to:%
\begin{equation*}
g\left( u^{\prime },u\right) =\cosh q^{\prime }\cosh q-\sinh q^{\prime
}\sinh q\cos \left( \chi ^{\prime }-\chi \right) .
\end{equation*}%
In the special case $\chi ^{\prime }=\chi $ we get the relative velocity:%
\begin{equation*}
v=\frac{\tanh \left( q^{\prime }-q\right) }{\sqrt{g_{0}}}\sinh q\partial
_{t}+\frac{\tanh \left( q^{\prime }-q\right) }{\sqrt{g_{1}}}\cosh q\cos \chi
\partial _{r}+\frac{\tanh \left( q^{\prime }-q\right) }{\sqrt{g_{3}}}\cosh
q\sin \chi \partial _{\phi }.
\end{equation*}%
One can see that this velocity is the space-like vector such that:%
\begin{equation}
v=u_{\left( 1\right) }\tanh \left( q^{\prime }-q\right)  \tag{2.44}
\end{equation}%
and $u_{\left( 1\right) }$ is given by eq. $\left( 2.6\right) $. For the
radial observers $\left( \chi ^{\prime }=\chi =0\right) $ with the constant
accelerations $\alpha $ and $\alpha ^{\prime }$ we obtain:%
\begin{equation*}
v^{t}=\frac{\sqrt{x^{2}-1}}{\sqrt{g_{0}}}\tanh \left[ \ln \left( \frac{y+%
\sqrt{y^{2}-1}}{x+\sqrt{x^{2}-1}}\right) \right] ,
\end{equation*}%
\begin{equation*}
v^{r}=\frac{x}{\sqrt{g_{1}}}\tanh \left[ \ln \left( \frac{y+\sqrt{y^{2}-1}}{%
x+\sqrt{x^{2}-1}}\right) \right] ,
\end{equation*}%
where:%
\begin{equation*}
x=\frac{1}{\sqrt{g_{0}}}\left( E+\alpha \int^{r}\sqrt{g_{0}g_{1}}dr\right)
\end{equation*}%
and%
\begin{equation*}
y=\frac{1}{\sqrt{g_{0}}}\left( E^{\prime }+\alpha ^{\prime }\int^{r}\sqrt{%
g_{0}g_{1}}dr\right) .
\end{equation*}

\section{Null-geodesics in the observers frames with the constant
acceleration}

Let us now consider the null-geodesics that correspond to the rays in the
geometric optics in the radial observer's frame. The null-geodesic is given
by the null vector field $k=k^{\mu }\partial _{\mu }$ which is tangent to a
curve $\gamma $ parametrized by an affine parameter $\lambda .$ Then the
equation of $\gamma $\ in the coordinate system $x^{\mu }$ is obtained from
the relations:%
\begin{equation}
\frac{\partial L}{\partial x^{\mu }}-\frac{d}{d\lambda }\left( \frac{%
\partial L}{\partial \overset{\cdot }{x}^{\mu }}\right) =0,  \tag{3.1}
\end{equation}%
where $L\left( x,\overset{\cdot }{x}\right) =\frac{1}{2}g_{\mu \nu }\left(
x\right) k^{\mu }k^{\nu }=0$ and $k^{\nu }\equiv \overset{\cdot }{x}^{\nu }$%
. For the metric (2.4) the eqs. (3.1) have the solutions: 
\begin{equation}
g_{0}k^{0}=e,\text{ \ }g_{3}k^{3}=j,  \tag{3.2}
\end{equation}%
where $e$ and $j$ can be interpreted as an energy and an angular momentum of
a photon, respectively. The constraint $L=0$ yields the relation:%
\begin{equation}
\left( \frac{dr}{d\lambda }\right) ^{2}=\left( k^{1}\right) ^{2}=\frac{1}{%
g_{1}}\left( \frac{e^{2}}{g_{0}}-\frac{j^{2}}{r^{2}}\right)  \tag{3.3}
\end{equation}%
The trajectories (3.1) are planar, so we fixed the coordinate $\theta =\pi
/2 $ which puts the constraint on a impact parameter $b=j/e$ of the photon: $%
r^{2}/g_{0}>b^{2}$ for $g_{0}\left( r\right) >0$. This equation can be
rewritten as follows:%
\begin{equation}
\left( \frac{dR}{jd\lambda }\right) ^{2}+V\left( R\right) =\frac{1}{b^{2}}, 
\tag{3.4}
\end{equation}%
where the new function $R=R\left( r\right) $ is given by the equation: $dR=%
\sqrt{g_{0}g_{1}}dr$ and $V\left( R\right) =g_{0}\left( r\left( R\right)
\right) /r^{2}\left( R\right) $. This equation has the standard form of a
particle motion in the potential $V\left( R\right) $. Thus the extremum of $%
V $ is obtained from the relation $dV/dR=0$ which leads to the equation:%
\begin{equation}
\frac{1}{r^{3}\sqrt{g_{0}\left( r\right) g_{1}\left( r\right) }}\left( r%
\frac{dg_{0}}{dr}-2g_{0}\left( r\right) \right) =0.  \tag{3.5}
\end{equation}%
The solutions of this equation we call $r_{e}^{\left( p\right) }:$%
\begin{equation}
\left. \frac{dV}{dR}\right\vert _{R_{e}=R\left( r_{e}^{\left( p\right)
}\right) }=0,  \tag{3.6}
\end{equation}%
where $p$ numbers of the consecutive solutions of (3.5). We need to know the
sign of $V^{\prime \prime }\left( R\left( r_{e}\right) \right) $.It will
depend on the form of the spherically symmetric metric which is the solution
of the Einstein equations with an energy-momentum tensor $T_{\mu \nu }$. It
is well known that if $T_{\mu \nu }=0$, then are only: the Minkowski or the
Schwarzschild-like metrics. Here we consider spherically symmetric metrics
with the non vanishing energy-momentum tensor. Thus we can not apriori
exclude the case where besides of the maximum of $V$ there are minima of $V.$
The maximum of $V$ we call $r_{M}$.and minimum $r_{m}$ Thus if $%
b^{-2}>V\left( R\left( r_{M}\right) \right) ,$ then the whole spacetime is
available for the trajectory of the photon (modulo horizons, if there are).
If $b^{-2}<V\left( R\left( r_{M}\right) \right) ,$ then the photon will hit
the potential barrier and return to the infinity with $dr/d\lambda >0$. In
such a case there would arise a region between $r_{M\text{ }}$ and $r_{m}$
where trapped photon trajectories would exist.

According to [6 (Box 25.7) ] $\psi $ is defined as an angle between
propagation direction of a photon\ and the radial direction of the observer.
In the case of the Schwarzschild metric a cone of avoidance with an apex in
an event $x_{0}^{\mu }$ is defined [7 (ch. 20) ] by the null critical
geodesics that passes $x_{0}^{\mu }$. An angle $\psi $ is an half-angle of
the cone.

In the observer frame with the tetrad $\left( u_{\left( \alpha \right)
}^{\mu }\right) $ the tangent vector field $k=k^{\mu }\partial _{\mu }$ to $%
\gamma $ has the components:%
\begin{equation}
k_{\left( \alpha \right) }=g_{\mu \nu }k^{\mu }u_{\left( \alpha \right)
}^{\nu }.  \tag{3.7}
\end{equation}%
Hence the four-vector $k$ is: $k=k_{\left( 0\right) }u+k_{\left( 1\right)
}u_{\left( 1\right) }+k_{\left( 3\right) }u_{\left( 3\right) }$ and $\left(
k_{\left( 0\right) }\right) ^{2}=\left( k_{\left( 1\right) }\right)
^{2}+\left( k_{\left( 3\right) }\right) ^{2}$. Thus $\psi $ is the angle
between the spacelike vector $\mathbf{k=}k_{\left( 1\right) }u_{\left(
1\right) }+k_{\left( 3\right) }u_{\left( 3\right) }$ and the axis $u_{\left(
3\right) }$ in the frame of the observer is expressed in arbitrary metric as
follows:%
\begin{equation}
\tan \psi =\frac{k_{\left( 1\right) }}{k_{\left( 3\right) }}.  \tag{3.8}
\end{equation}%
In the spherically symmetric space-time with the diagonal metric and a
radially moving observer we get:%
\begin{equation}
k_{\left( 0\right) }=\frac{e}{\sqrt{g_{0}}}\cosh q\pm \sqrt{\frac{e^{2}}{%
g_{0}}-\frac{j^{2}}{g_{3}}}\sinh q,  \tag{3.9}
\end{equation}%
\begin{equation}
k_{\left( 1\right) }=\frac{e}{\sqrt{g_{0}}}\sinh q\pm \sqrt{\frac{e^{2}}{%
g_{0}}-\frac{j^{2}}{g_{3}}}\cosh q,  \tag{3.10}
\end{equation}%
\begin{equation}
k_{\left( 3\right) }=\frac{j}{\sqrt{g_{3}}}.  \tag{3.11}
\end{equation}%
The $-$ sign corresponds to photons coming from the infinity to the observer
and $+$ sign corresponds to the photons that can reach the observer provided
their impact parameter satisfies: $b^{-2}<V\left( R\left( r_{M}\right)
\right) $ \ It leads to the relation:%
\begin{equation}
\tan \psi _{\pm }=\frac{1}{b}\sqrt{\frac{g_{3}}{g_{0}}}\left[ \sinh q\pm 
\sqrt{1-b^{2}\frac{g_{0}}{g_{3}}}\cosh q\right] ,  \tag{3.12}
\end{equation}
One can rewrite the last formula as follows:%
\begin{equation}
\tan \psi _{\pm }=\sinh \left( q\pm w\right) ,  \tag{3.12a}
\end{equation}%
where $\cosh w=\sqrt{g_{3}/\left( b^{2}g_{0}\right) }$. For the radial
observer with the constant acceleration (2.30) and $\theta =\pi /2$ ($%
g_{3}=r^{2}$) we obtain:%
\begin{equation}
q\pm w=\ln \left[ \left\vert b\right\vert \frac{s+\sqrt{s^{2}-g_{0}}}{r\mp 
\sqrt{r^{2}-b^{2}g_{0}}}\right] ,  \tag{3.13}
\end{equation}%
where $s=E+\alpha \int^{r}\sqrt{g_{0}g_{1}}dr$. Hence $q+w\rightarrow \infty 
$ if $g_{0}\rightarrow 0$ what means that $\psi \rightarrow \pi /2$ on the
horizon.

\subsection{Reissner--Nordstr\"{o}m--de Sitter spacetime}

This spacetime is a spherically symmetric solution of the Einstein equations
[8] described by the three parameters: a mass $m$, an electric charge $e$
and a cosmological constant $\Lambda $ ; it generalizes the Schwarzschild
spacetime. It represents a black hole, that generally possess three horizons
according to the nonzero values of $m,$ $e$ and $\Lambda $. These horizons
correspond to outer (event) and inner (Cauchy) black hole horizons and the
cosmological horizon. It may therefore be interpreted as the space-time of a
charged black hole in a de Sitter or anti-de Sitter spacetimes. The metric
is:%
\begin{equation}
ds^{2}=g_{0}\left( r\right) dt^{2}-\frac{1}{g_{0}\left( r\right) }%
dr^{2}-r^{2}\left( d\theta ^{2}+\sin ^{2}\theta d\phi ^{2}\right) , 
\tag{3.14}
\end{equation}%
where:%
\begin{equation*}
g_{0}\left( r\right) =1-\frac{2m}{r}+\frac{e^{2}}{r^{2}}-\frac{\Lambda }{3}%
r^{2}.
\end{equation*}%
In order to simplify calculations we consider the uncharged black hole $e=0$
in the de Sitter $\Lambda >0$ or anti-deSitter $\Lambda <0$ spacetimes. Thus
the horizons are given by the solution of the cubic equation: 
\begin{equation*}
r^{3}-\frac{3}{\Lambda }r+\frac{6m}{\Lambda }=0
\end{equation*}

In the de Sitter case and $m>0$ one gets that the number of real roots
depends on the sign of the discriminant $D=-1/\Lambda ^{3}+9m^{2}/\Lambda
^{2}$ . If $D>0$ (it means that: $9m^{2}\Lambda >1$), then there is no real
positive root. If $D<0$ (it means that: $0<9m^{2}\Lambda <1$), then there
are two real positive roots $r_{+}>r_{-}>0$:%
\begin{equation}
r_{\pm }=\frac{2}{\sqrt{\Lambda }}\cos \left[ \frac{\pi }{3}\mp \frac{1}{3}%
\arccos \left( 3m\sqrt{\Lambda }\right) \right]  \tag{3.15}
\end{equation}%
and one negative root $r_{0}<0.$ Moreover the sum of these roots vanishes: $%
r_{+}+r_{-}+r_{0}=0.$ Hence one gets:%
\begin{equation*}
r_{0}=-\frac{2}{\sqrt{\Lambda }}\cos \left[ \frac{1}{3}\arccos \left( 3m%
\sqrt{\Lambda }\right) \right] .
\end{equation*}%
These positive roots determine the cosmological horizon $\left( r_{+}\right) 
$ and the event horizon $\left( r_{-}\right) $. Hence the metric component $%
g_{0}$ takes the form:%
\begin{equation*}
g_{0}\left( r\right) =\frac{\Lambda }{3r}\left( r_{+}-r\right) \left(
r-r_{-}\right) \left( r-r_{0}\right)
\end{equation*}%
and the metric is stationary for $r\in \left( r_{-},r_{+}\right) $. One can
notice that for $9m^{2}\Lambda =1$ the cosmological and event horizons
coincide: $r_{+}=r_{-}=3m=1/\sqrt{\Lambda }$ and:%
\begin{equation*}
g_{0}\left( r\right) =-\frac{\Lambda }{3r}\left( r-3m\right) ^{2}\left(
r+6m\right) .
\end{equation*}%
In this case the coordinate $r$ becomes temporal-like while $t$ becomes
spatial-like. If $m=0,$ then the metric is the de Sitter metric in the
static coordinates and the cosmological horizon is given by: $r_{c}=\sqrt{%
3/\Lambda }$.

In the anti-de Sitter case the equation $g_{0}\left( r\right) $ has the one
real root:%
\begin{equation}
r_{h}=\frac{2}{\sqrt{-\Lambda }}\sinh \left[ \frac{1}{3}\sinh ^{-1}\left( 3m%
\sqrt{-\Lambda }\right) \right] ,  \tag{3.16}
\end{equation}%
where $\sinh ^{-1}\left( y\right) $ is the inverse of the $\sinh x=y$.

The angle $\psi $ for $\Lambda >0$ is given by:%
\begin{equation}
\tan \psi =\frac{r}{b}\sqrt{\frac{3r}{\Lambda f\left( r\right) }}\left[
\sinh q\pm \sqrt{1-b^{2}\frac{\Lambda f\left( r\right) }{3r^{3}}}\cosh q%
\right] ,  \tag{3.17}
\end{equation}%
where: $f\left( r\right) =\left( r_{+}-r\right) \left( r-r_{-}\right) \left(
r+r_{0}\right) =-r^{3}+3r/\Lambda -6m/\Lambda $.

In the case of the Schwarzschild metric the eq. (3.17) is:%
\begin{equation}
\tan \psi =\frac{1}{\widetilde{b}\sqrt{y-1}}\left[ \frac{1}{\sqrt{y}}\sinh
q\pm \sqrt{y^{3}-\widetilde{b}^{2}y+\widetilde{b}^{2}}\cosh q\right] , 
\tag{3.18}
\end{equation}%
where we introduced the dimensionless variables $y=r/r_{S}>1$ and $%
\widetilde{b}=b/r_{S}$. Hence the psi angle is function of two parameters: $%
q $ and $\widetilde{b}$. For the observer with the constant acceleration
eqs.(2.30-31) we obtain the formula:%
\begin{eqnarray}
\tan \psi &=&\frac{\sqrt{y}}{\widetilde{b}\left( y-1\right) }\left[ \frac{1}{%
y}\sqrt{y\left( E-r_{s}\alpha +yr_{s}\alpha \right) ^{2}+1-y}\right.  \notag
\\
&&\left. \pm \sqrt{y^{3}-\widetilde{b}^{2}y+\widetilde{b}^{2}}\left(
E-r_{s}\alpha +yr_{s}\alpha \right) \right] .  \TCItag{3.19}
\end{eqnarray}%
For the geodesic radial observer ($\alpha =0)$ the above formula reads:%
\begin{equation}
\tan \psi =\frac{\sqrt{y}}{\widetilde{b}\left( y-1\right) }\left[ \frac{1}{y}%
\sqrt{y\left( E^{2}-1\right) +1}\pm E\sqrt{y^{3}-\widetilde{b}^{2}y+%
\widetilde{b}^{2}}\right] .  \tag{3.20}
\end{equation}

For the static observer ($\sinh q=0$) the formula (3.18) gives the relation:%
\begin{equation}
\cot ^{2}\psi \left( y;\widetilde{b}\right) =\widetilde{b}^{2}\frac{y-1}{%
y^{3}-\widetilde{b}^{2}y+\widetilde{b}^{2}}.  \tag{3.21}
\end{equation}%
The other form of (3.19) is:%
\begin{equation}
0\leq \sin ^{2}\psi \left( y;\widetilde{b}\right) =\widetilde{b}^{2}\frac{y-1%
}{y^{3}}\leq 1.  \tag{3.22}
\end{equation}%
From the last equation we get the constraint relating $\widetilde{b}$ and $%
y: $%
\begin{equation}
y^{3}-\widetilde{b}^{2}y+\widetilde{b}^{2}\geq 0.  \tag{3.23}
\end{equation}

\section{Observers in the spacetime with the axial symmetry}

In this section we will consider a spacetime with the axial symmetry having
two Killing vectors $V_{\left( t\right) }=\partial _{t}$ and $V_{\left( \phi
\right) }=\partial _{\phi }$. A metric $g_{\mu \nu }$ has the non-diagonal
form:%
\begin{equation}
ds^{2}=g_{0}dt^{2}-g_{1}dr^{2}-g_{2}d\theta ^{2}-g_{3}d\phi ^{2}-2hdtd\phi 
\tag{4.1}
\end{equation}%
and the coefficients $g_{a}$.and $h$ are the functions of $r$ and $\theta $.
An observer in the space-time (4.1) for $\theta =const$ is given by a
time-like vector field $u$:%
\begin{equation}
u\equiv u_{\left( 0\right) }=\alpha \partial _{t}+\beta \partial _{r}+\gamma
\partial _{\phi }.  \tag{4.2}
\end{equation}%
The condition $u^{2}=1$ takes the form: $\alpha ^{2}D/g_{3}-\beta
^{2}g_{1}-\left( \alpha h+\gamma g_{3}\right) ^{2}/g_{3}=1,$ where: $%
D=h^{2}+g_{0}g_{3}$ \ Also in this case we can parametrize the vector field $%
u$ by a pair of functions: $\left( q\left( r\right) ,\chi \left( r\right)
\right) $:%
\begin{equation}
u=\sqrt{\frac{g_{3}}{D}}\cosh q\partial _{t}+\frac{1}{\sqrt{g_{1}}}\sinh
q\cos \chi \partial _{r}+\frac{1}{\sqrt{g_{3}}}\left( \sinh q\sin \chi -%
\frac{h}{\sqrt{D}}\cosh q\right) \partial _{\phi }.  \tag{4.3}
\end{equation}%
The other three basis vectors $u_{\left( i\right) }$ are:%
\begin{equation}
u_{\left( 1\right) }=\sqrt{\frac{g_{3}}{D}}\sinh q\partial _{t}+\frac{1}{%
\sqrt{g_{1}}}\cosh q\cos \chi \partial _{r}+\frac{1}{\sqrt{g_{3}}}\left(
\cosh q\sin \chi -\frac{h}{\sqrt{D}}\sinh q\right) \partial _{\phi }, 
\tag{4.4}
\end{equation}%
\begin{equation}
u_{\left( 2\right) }=\frac{1}{\sqrt{g_{2}}}\partial _{\theta },  \tag{4.5}
\end{equation}%
\begin{equation}
u_{\left( 3\right) }=-\frac{\sin \chi }{\sqrt{g_{1}}}\partial _{r}+\frac{%
\cos \chi }{\sqrt{g_{3}}}\partial _{\phi }.  \tag{4.6}
\end{equation}%
In this way we determine the observer whose frame of reference is given by
(4.3-6). In the case $h=0$ we get (2.5-8). The integral curve for (4.3) is
given by the equations:%
\begin{equation}
\frac{dt}{d\lambda }=\sqrt{\frac{g_{3}}{D}}\cosh q,  \tag{4.7}
\end{equation}%
\begin{equation}
\frac{dr}{d\lambda }=\frac{1}{\sqrt{g_{1}}}\sinh q\cos \chi ,\text{ \ }\frac{%
d\theta }{d\lambda }=0,  \tag{4.8}
\end{equation}%
\begin{equation}
\frac{d\phi }{d\lambda }=\frac{1}{\sqrt{g_{3}}}\left( \sinh q\sin \chi -%
\frac{h}{\sqrt{D}}\cosh q\right) .  \tag{4.9}
\end{equation}%
From the equations (4.7-9) we can see that the static observer ($dr/d\lambda
=d\theta /d\lambda =d\phi /d\lambda =0$) is realized by the functions: $\chi
=\pi /2$ and $\tanh q=h/\sqrt{D}$ so $\cosh q=\sqrt{D/\left(
g_{0}g_{3}\right) }$ and $\sinh q=h/\sqrt{g_{0}g_{3}}$. The radial
trajectories $\left( \phi =const\right) $ are given by the conditions 
\begin{equation}
\tanh q\sin \chi =h/\sqrt{D}\text{ \ and }\chi \neq 0.  \tag{4.10}
\end{equation}

Next we consider a null-geodesic with a tangent vector $k=k^{\mu }\partial
_{\mu }$ in this space-time. The Euler-Lagrange equations gives the
conserved quantities $e$ and $j$:%
\begin{equation}
g_{0}k^{0}-hk^{3}=e,  \tag{4.11}
\end{equation}%
\begin{equation}
hk^{0}+g_{3}k^{3}=j.  \tag{4.12}
\end{equation}%
It is easy solve the above equations so:%
\begin{equation}
k^{0}=\frac{D_{0}\left( e,j\right) }{D},\text{ \ }k^{3}=\frac{D_{3}\left(
e,j\right) }{D},  \tag{4.13}
\end{equation}%
where $D_{0}\left( e,j\right) =hj+g_{3}e$ and $D_{3}\left( e,j\right)
=-he+g_{0}j$. There is the constraint $k\cdot k=0$ that provides the
relation between the $k^{1}$ and $k^{2}$:%
\begin{equation}
g_{1}\left( k^{1}\right) ^{2}+g_{2}\left( k^{2}\right) ^{2}=\frac{W\left(
r,\theta ;e,j\right) }{g_{3}D},  \tag{4.14}
\end{equation}%
where: $W\left( r,\theta ;e,j\right) \equiv D_{0}^{2}-j^{2}D$. For the Kerr
metric the constraint (4.14) can be separated. It means that there are two
functions $R\left( r\right) $ and $\Theta \left( \theta \right) $ such that: 
$k^{1}/\sqrt{R\left( r\right) }=\pm k^{2}/\sqrt{\Theta \left( \theta \right) 
}$ (see Section 5). We can solve the constraint (4.14) introducing a new
function $\sigma $ of $r$ and $\theta $ such that:%
\begin{equation}
k^{1}=\sqrt{\frac{W}{g_{1}g_{3}D}}\cos \sigma \left( r,\theta \right) ,\text{
\ }k^{2}=\sqrt{\frac{W}{g_{2}g_{3}D}}\sin \sigma \left( r,\theta \right) . 
\tag{4.15}
\end{equation}%
The function $\sigma $ has the interpretation of an angle in the plane $%
\left( k^{1},k^{2}\right) $.

If the null vector $k$ has the second component $k^{2}$ equal to zero, then $%
\psi $ in the observer's frame is given by the eq. (3.8). One can say that
the null geodesic given by $k$ is confined to the equatorial plane $\theta
=\pi /2$ and $k^{1}=\sqrt{W/\left( g_{1}g_{3}D\right) }$. Then the equation
(3.7) gives the components of $k$ in the observer's frame:%
\begin{equation}
k_{\left( 1\right) }=\frac{D_{0}}{\sqrt{g_{3}D}}\sinh q-\frac{1}{\sqrt{g_{3}}%
}\left( \sqrt{W/D}\cos \chi +j\sin \chi \right) \cosh q,  \tag{4.16}
\end{equation}%
\begin{equation}
k_{\left( 2\right) }=0\text{, \ \ }k_{\left( 3\right) }=\sqrt{\frac{W}{g_{3}D%
}}\sin \chi -\frac{j}{\sqrt{g_{3}}}\cos \chi .  \tag{4.17}
\end{equation}%
Hence we get:%
\begin{equation}
\tan \psi \left( r;q,\chi \right) =\frac{1}{\sqrt{D}}\frac{D_{0}\sinh
q-\left( \sqrt{\frac{D_{0}^{2}}{D}-j^{2}}\cos \chi +j\sin \chi \right) \sqrt{%
D}\cosh q}{\sqrt{\frac{D_{0}^{2}}{D}-j^{2}}\sin \chi -j\cos \chi }, 
\tag{4.18}
\end{equation}%
where $D_{0}=D_{0}\left( e,j\right) $ depends on the parameters $e$ and $j$
describing the null-geodesic.

For the radial trajectory of the observer in the equatorial plane ($\tanh
q\sin \chi =h/\sqrt{D}$) and for $\chi =\pi /4$ there is the relation: $%
\sqrt{D}\sinh q=h\sqrt{2}\cosh q$ so the observer velocity vector $u$ takes
the form:%
\begin{equation}
u=\sqrt{\frac{g_{3}}{D}}\cosh q\partial _{t}+\frac{h}{\sqrt{g_{1}D}}\cosh
q\partial _{r}.  \tag{4.19}
\end{equation}%
Moreover from the condition $u^{2}=+1$ we get: $\cosh
q=D^{1/2}(g_{0}g_{3}-h^{2})^{-1/2}$. Thus the formula (4.18) is:%
\begin{equation}
\tan \psi =\frac{1}{\sqrt{D(g_{0}g_{3}-h^{2})}}\times \frac{2D_{0}h-\left( 
\sqrt{\frac{D_{0}^{2}}{D}-j^{2}}+j\right) D}{\sqrt{\frac{D_{0}^{2}}{D}-j^{2}}%
-j}.  \tag{4.20}
\end{equation}

\section{Kerr metric}

In this section we will apply the results of the section 4 to the Kerr
metric. This metric in terms of the spheroidal-like coordinates of Boyer and
Lindquist (1967) has the form:%
\begin{eqnarray}
ds^{2} &=&\left( 1-\frac{2mr}{\Sigma }\right) dt^{2}+4\frac{mar\sin
^{2}\theta }{\Sigma }dtd\phi -\frac{\Sigma }{\Delta }dr^{2}  \notag \\
&&-\Sigma d\theta ^{2}-\left( r^{2}+a^{2}+2\frac{mra^{2}\sin ^{2}\theta }{%
\Sigma }\right) \sin ^{2}\theta d\phi ^{2}.  \TCItag{5.1}
\end{eqnarray}%
where $\Delta \left( r\right) =r^{2}-2mr+a^{2}$ and $\Sigma \left( r,\theta
\right) =r^{2}+a^{2}\cos ^{2}\theta .$ The variables $\left( t,r,\theta
,\phi \right) $ belong to the following intervals: $t\in R^{1},$ $r\in
\left( 0,r_{-}\right) \cup \left( r_{-},r_{+}\right) \cup \left(
r_{+},+\infty \right) ,$ $\theta \in \left( 0,\pi \right) $ and $\phi \in
\left( 0,2\pi \right) $ where: $r_{\pm }=m\pm \sqrt{m^{2}-a^{2}}$.

In order to obtain the angle \ $\psi \ $(4.18) we have to know the component 
$k^{1}$ and $D_{0}\left( e,j\right) $ of the null vector field $k$ and the
function $q$ and $\chi $ defining the observer. The components $k^{\mu }$ of
the null geodesic vector field $k=k^{\mu }\partial _{\mu }$ are well-known
(e.g [9] ) and have the form: 
\begin{equation*}
k^{0}=\frac{1}{\Sigma }\left[ a\left( j-ae\sin ^{2}\theta \right) +\frac{%
r^{2}+a^{2}}{\Delta }P\right] ,
\end{equation*}

\begin{equation*}
k^{1}=\frac{\sqrt{R}}{\Sigma },\text{ \ }k^{2}=\frac{\sqrt{\Theta }}{\Sigma }%
,
\end{equation*}%
\begin{equation*}
k^{3}=-\frac{1}{\Sigma }\left[ \left( ae-\frac{j}{\sin ^{2}\theta }\right) +%
\frac{a}{\Delta }P\right] ,
\end{equation*}%
where:%
\begin{equation*}
P\left( r\right) =e\left( r^{2}+a^{2}\right) -ja,
\end{equation*}%
\begin{equation*}
R\left( r\right) =P^{2}-\left[ C+\left( j-ae\right) ^{2}\right] \Delta ,
\end{equation*}%
\begin{equation*}
\Theta \left( \theta \right) =C+a^{2}\left( e^{2}-\frac{j^{2}}{a^{2}\sin
^{2}\theta }\right) \cos ^{2}\theta
\end{equation*}%
and $C$ is the separation constant in the equation (4.14). These components
are the solutions of the general algebraic equations (4.13-14) for the axial
symmetric spacetime. Hence we get:%
\begin{equation}
D=\Delta \left( r\right) \sin ^{2}\theta ,  \tag{5.2}
\end{equation}%
\begin{equation}
D_{0}=\left[ e\left( r^{2}+a^{2}\right) \Sigma -2mra\left( j-a\sin
^{2}\theta \right) \right] \frac{\sin ^{2}\theta }{\Sigma },  \tag{5.3}
\end{equation}%
\begin{equation}
D_{3}=j\left( 1-\frac{2mr}{\Sigma }\right) +2e\frac{mar\sin ^{2}\theta }{%
\Sigma }.  \tag{5.4}
\end{equation}%
We want to find the psi angle given by (4.18) so we only consider the null
vectors in the equatorial plane $\theta =\pi /2.$ Thus $\Sigma \left( r,\pi
/2\right) =r^{2},$ $g_{3}\left( \pi /2\right) =r^{2}+a^{2}+2ma^{2}/r,$ $%
D\left( \pi /2\right) =\Delta \left( r\right) $ and:

\begin{equation*}
D_{0}\left( \pi /2\right) =\left[ e\left( r^{2}+a^{2}\right) r-2ma\left(
j-a\right) \right] \frac{1}{r}.
\end{equation*}

We obtain then\bigskip

\begin{equation}
\tan \psi \left( r;q,\chi \right) =\frac{D_{0}\sinh q-\left( \sqrt{%
D_{0}^{2}-j^{2}\Delta }\cos \chi +j\sqrt{\Delta }\sin \chi \right) \cosh q}{%
\sqrt{D_{0}^{2}-j^{2}\Delta }\sin \chi -j\sqrt{\Delta }\cos \chi }  \tag{5.5}
\end{equation}%
\ 

For the radial motion and $\chi =\pi /4$ \ we get\ from (4.19) :%
\begin{equation}
\tan \psi =\frac{1}{\sqrt{\Delta (g_{0}g_{3}-h^{2})}}\times \frac{%
2D_{0}h-\left( \sqrt{\frac{D_{0}^{2}}{\Delta }-j^{2}}+j\right) \Delta }{%
\sqrt{\frac{D_{0}^{2}}{\Delta }-j^{2}}-j},  \tag{5.6}
\end{equation}%
where $h(r)=2ma/r$ and $g_{0}=1-2m/r$. Thus the angle $\psi $ depends on the
two parameters of the null-geodesic: $e$, $j$.

In the case of the geodesic observer in the equatorial plane $\left( d\theta
/d\lambda =0\right) $ the equations (4.7-9) are known in the explicit form:%
\begin{equation*}
\overset{\cdot }{t}\equiv \frac{dt}{d\lambda }=\frac{1}{\Delta }\left[
\left( r^{2}+a^{2}+\frac{2ma^{2}}{r}\right) E-\frac{2ma}{r}J\right] ,
\end{equation*}%
\begin{equation*}
\overset{\cdot }{r}^{2}\equiv \left( \frac{dr}{d\lambda }\right) ^{2}=\frac{1%
}{r^{2}}\left[ -\Delta +r^{2}E^{2}+\frac{2m}{r}\left( J-aE\right)
^{2}-\left( J^{2}-a^{2}E^{2}\right) \right] ,
\end{equation*}%
\begin{equation*}
\overset{\cdot }{\phi }\equiv \frac{d\phi }{d\lambda }=\frac{1}{\Delta }%
\left[ \left( 1-\frac{2m}{r}\right) J+\frac{2ma}{r}E\right] ,
\end{equation*}%
where $E$ and $J$ are an energy and an angular momentum of the observer,
respectively. From these equations one gets the functions $q$ and $\chi $
describing the observer. Next inserting functions $q$ and $\chi $ into the
eq. (5.5) we get the following formula for the angle $\psi $:%
\begin{equation*}
\tan \psi =\frac{1}{\sqrt{\Delta }}\frac{D_{0}\left( \overset{\cdot }{t}%
^{2}g_{3}-\Delta \right) -\sqrt{\Delta }\left( \overset{\cdot }{r}\sqrt{g_{1}%
}\sqrt{D_{0}^{2}-j^{2}\Delta }+j\sqrt{\overset{\cdot }{t}^{2}g_{3}-\Delta
\left( 1+\overset{\cdot }{r}^{2}g_{1}\right) }\right) }{\sqrt{%
D_{0}^{2}-j^{2}\Delta }\sqrt{\overset{\cdot }{t}^{2}g_{3}-\Delta \left( 1+%
\overset{\cdot }{r}^{2}g_{1}\right) }-j\overset{\cdot }{r}\sqrt{g_{1}}},
\end{equation*}%
where $g_{1}=\left( 1-2m/r+a^{2}/r^{2}\right) ^{-1}$.

For the slow rotating Kerr black hole $a^{2}/r^{2}<<1$ the metric takes the
form:%
\begin{equation*}
ds^{2}=\left( 1-\frac{2m}{r}\right) dt^{2}+4\frac{ma\sin ^{2}\theta }{r}%
dtd\phi -\left( 1-\frac{2m}{r}\right) ^{-1}dr^{2}-r^{2}d\Omega ^{2},
\end{equation*}%
where $d\Omega ^{2}=d\theta ^{2}+\sin ^{2}\theta d\phi ^{2}$. In this case
there is only one horizon: $r_{h}=2m$. In the eq. (5.5) survived only terms
linear in $a.$ Thus $\psi $ is given by:%
\begin{equation*}
\tan \psi \left( r;q,\chi \right) =\frac{D_{0}\sinh q-\left( \sqrt{F}\cos
\chi +j\sqrt{\Delta }\sin \chi \right) \cosh q}{\sqrt{F}\sin \chi -j\sqrt{%
\Delta }\cos \chi },
\end{equation*}%
where $F=e^{2}r^{4}-j^{2}r^{2}+2mrj\left( j-2ae\right) $, $%
D_{0}=er^{2}-2maj/r$ and $\Delta =r^{2}-2mr$.

In the general case we can obtain the pair $\left( q,\chi \right) $ and
inserting them into the general formula (4.18) one obtains $\tan \psi $. As
one can see this result will be highly complicated and crucially depending
on the three null geodesic\ parameters $\left( e,j,C\right) $ and three
observer's parameters $\left( E,J,K\right) $ . Here parameter $K$ is the
separation constant of the observer's time-like geodesic.

\section{Conclusions}

We have obtained the family of observers parametrized by the pair of
functions $\left( q,\chi \right) $. This parametrization was applied to
determine the angle $\psi $ in the spacetimes with spherical and axial
symmetries. The angle $\psi $ depends on the parameters describing the null
geodesic of the photon $\left( e,j\right) $ and parameters $\left( q,\chi
\right) $\ describing an observer. In the case of the spherical symmetry we
get the family of the observers with the constant acceleration. For the
radial and transversal observers we obtained the functions $q$ in the
explicit form (2.30) and (2.39) respectively. For the radial observer with
the constant acceleration and the Schwarzschild metric we obtained the
equation (3.11) for the the psi angle.

We extended our considerations to the more realistic spacetimes with the
axial symmetry in the sections 4 and 5. In the case of the Kerr metric we
obtained the compact expression for $\psi $ in the case of the radial motion.

The obtained parametrization can be used in other applications.

\section{Appendix}

The non-vanishing Christoffel symbols for the metric (2.4) are:%
\begin{eqnarray*}
\Gamma _{tr}^{t} &=&\frac{g_{0}^{\prime }}{2g_{0}},\text{ \ } \\
\Gamma _{tt}^{r} &=&\frac{g_{0}^{\prime }}{2g_{1}},\text{ \ }\Gamma
_{rr}^{r}=\frac{g_{1}^{\prime }}{2g_{1}},\text{ \ }\Gamma _{\theta \theta
}^{r}=-\frac{r}{g_{1}},\text{ \ }\Gamma _{\phi \phi }^{r}=-\frac{r\sin
^{2}\theta }{g_{1}}, \\
\Gamma _{r\theta }^{\theta } &=&\frac{1}{r},\text{ \ }\Gamma _{\phi \phi
}^{\theta }=-\sin \theta \cos \theta , \\
\Gamma _{r\phi }^{\phi } &=&\frac{1}{r},\text{ \ }\Gamma _{\theta \phi
}^{\phi }=\cot \theta .
\end{eqnarray*}%
The acceleration $a$ of the vector field (2.5) for the metric (2.4) is:%
\begin{equation*}
a^{t}=2\Gamma _{tr}^{t}u^{r}u^{t}+u^{r}\partial _{r}u^{t},
\end{equation*}%
\begin{equation*}
a^{r}=\Gamma _{tt}^{r}\left( u^{t}\right) ^{2}+\Gamma _{rr}^{r}\left(
u^{r}\right) ^{2}+\Gamma _{\phi \phi }^{r}\left( u^{\phi }\right)
^{2}+u^{r}\partial _{r}u^{r},
\end{equation*}%
\begin{equation*}
a^{\phi }=2\Gamma _{r\phi }^{\phi }u^{r}u^{\phi }+u^{r}\partial _{r}u^{\phi
},\text{ \ \ }a^{\theta }=0,
\end{equation*}%
where $u^{t}=\cosh q/\sqrt{g_{0}},$ $u^{r}=\sinh q\cos \chi /\sqrt{g_{1}}$
and $u^{\phi }=\sinh q\sin \chi /\sqrt{g_{3}}$.

\section{References}

[1] E. Cartan: \textit{Les probl\`{e}mes d'\'{e}quivalence}, Oeuvres Compl%
\`{e}tes, Volume 2, Issue 2, pg. 1311-1334

[2] C Rovelli, \textit{What is observable in classical and quantum gravity?}%
, Class Quant Grav 8 (1991) 297.

[3] H. Westman, S. Sonego: \textit{Coordinates, observables and symmetry in
relativity,}\ Annals of Physics 324 (2009) 1585-1611, [arXiv:0711.2651]\ 

[4] P. Duch, W. Kami\'{n}ski, J. Lewandowski, J. \'{S}wie\.{z}ewski: \textit{%
Observables for General Relativity related to geometry,} JHEP05(2014)077
[arXiv:1403.8062]

[5] V. Bolos: \textit{Intrinsic definitions of "relative velocity" in
general relativity,} Commun.Math.Phys.273:217-236,2007, [arXiv:
gr-qc/0506032]\ \ 

[6] C. W. Misner, K. S. Thorne, J. A. Wheeler: \textit{Gravitation}, San
Francisco 1973.

[7] S. Chandrasekhar: \textit{The Mathematical Theory of Black Holes},
Oxford 1983

[8] J. Griffiths, J. Podolsky:\ \textit{Exact Space-Times in Einstein's
General Relativity, }Cambridge University Press, 2009

[9] B. O'Neill: \textit{The Geometry of Kerr Black Holes}, Wellesley 1995

\end{document}